# Analytical Coexistence Benchmark for Assessing the Utmost Interference Tolerated by IEEE 802.20

Mouhamed Abdulla* and Yousef R. Shayan*

**Abstract**—Whether it is crosstalk, harmonics, or in-band operation of wireless technologies, interference between a reference system and a host of offenders is virtually unavoidable. In past contributions, a benchmark has been established and considered for coexistence analysis with a number of technologies including FWA, UMTS, and WiMAX. However, the previously presented model does not take into account the mobility factor of the reference node in addition to a number of interdependent requirements regarding the link direction, channel state, data rate and system factors; hence limiting its applicability for the MBWA (IEEE 802.20) standard. Thus, over diverse modes, in this correspondence we analytically derived the greatest aggregate interference level tolerated for high-fidelity transmission tailored specifically for the MBWA standard. Our results, in the form of benchmark indicators, should be of particular interest to peers analyzing and researching RF coexistence scenarios with this new protocol.

## 1. INTRODUCTION

Gradually, 4G systems are expected to dominate the marketplace in the years to come. In principle, there seems to be two general options available as we move toward this: either upgrade 3 and 3.5G or simply use a new technology. On the surface, some may consider an upgrade as a cost-effective solution because it requires minor infrastructure modifications or none at all. However, an upgrade would always be constraint to backward compatibility issues, and this would result in a suboptimal system which often defeats the purpose. Thus, due to this dominant reason, among other factors, a new cellular standard called mobile broadband wireless access (MBWA) or IEEE 802.20 was approved by the IEEE Standard Association Board [1].

Ultimately, the goal of this paper is to obtain a proper comparison benchmark in order to assess or quantify how different network configurations and concentrations of wireless interferers can impact an 802.20 node in coexistence-based research. Granted, such limitations have already been determined for a number of wireless systems such as: fixed wireless access (FWA), universal mobile telecommunications system (UMTS), and worldwide interoperability for microwave access (WiMAX - IEEE 802.16-2004) [2-4]. However, the premise considered for these analytical formulations does not take into consideration the dynamic factor of the reference and the impact that it may carry regarding a host of entangled requirements along with the link direction, channel access, and throughput. Therefore, a detailed and elaborate derivation

**Corresponding Author: Mouhamed Abdulla**
* Dept. of Electrical and Computer Engineering, Concordia University, Montréal, Québec, Canada (ma14@ieee.org, yshayan@ece.concordia.ca)

oriented exclusively for the MBWA standard is essential in anticipation to its deployment.

Specifically, in Section II of this contribution, we will briefly identify the major and relevant highlights of this new standard. Then, in Section III a careful description of the system model will be given. Next, in Section IV we will analytically derive a practical and appropriate margin for the maximum aggregate interference permitted by an MBWA portable device. After, in Section V numerical interpretation of the findings will be shown in order to clearly characterize the limitations of this novel technology. Finally, Section VI closes with noteworthy observations drawn from the presented treatment.

## 2. MBWA FEATURES

IEEE 802.16e-2005 has a data rate that could practically reach 10 Mbps over 2 km under no line of sight (NLOS); but can only support radios with vehicular speed of 60+ km/h [5]. On the other hand, currently operable cellular systems, irrespective of whether they are founded on the global system for mobile communications (GSM) or code division multiple access (CDMA), offer substantially higher mobility at the cost of a mediocre bandwidth. Therefore, it was natural to combine these advantages to form the essence of the 802.20 technology. In effect, when compared to other mobile systems, such as: enhanced data rates for GSM evolution (EDGE), UMTS, CDMA2000 1xRTT and 1xEV; MBWA has the highest spectral efficiency with an increased mobility of up to 250 km/h [6]. Hence, this protocol can be seen as the missing component between available WMAN and WWAN standards.

Further, the 802.20 specification only defines the lower PHY and MAC layers of the open systems interconnection (OSI) model; thus granting vast compatibility with an array of systems through the upper network levels. Also, it has low latency with a frame round trip time of at most 10 ms [6]. And in fact, there is a direct relation between latency and performance which may be traded among each other [7] to enhance real-time applications and to satisfy the service delivery. Overall, IEEE 802.20 is a technology that has many benefits; nonetheless, because it is a novel system, base-station infrastructure cost is inevitable.

## 3. SYSTEM MODEL

Cleary, MBWA is a cellular standard where a mobile is connected to a serving base station (BS) while active in its coverage area. Although interference at the BS is vital, in this paper we will only focus on analysis pertaining to the mobile end for both downlink (DL) and uplink (UL) transmissions.

Broadly speaking, the MBWA mobile receiver is composed of three fundamental elements, namely, the antenna, amplifier, and detector; where the received signal is captured, enhanced and processed into binary bits as depicted in Fig. 1. In particular, the signal constitutes the following components: the information wave $r(t)$ forwarded from the BS, a number of interferers $i_j(t)$ where $j$ represents the index for unwanted emissions for $\forall m \in \mathbb{N}^* : \exists j \in \mathbb{N}^* : 1 \le j \le m$, the inherent thermal noise due to the terminal circuitry of the antenna $n_{ant}(t)$, and the amplifier $n_{amp}(t)$; where all these random processes are defined over $\mathbb{R}^1_+ \mapsto \mathbb{R}$.

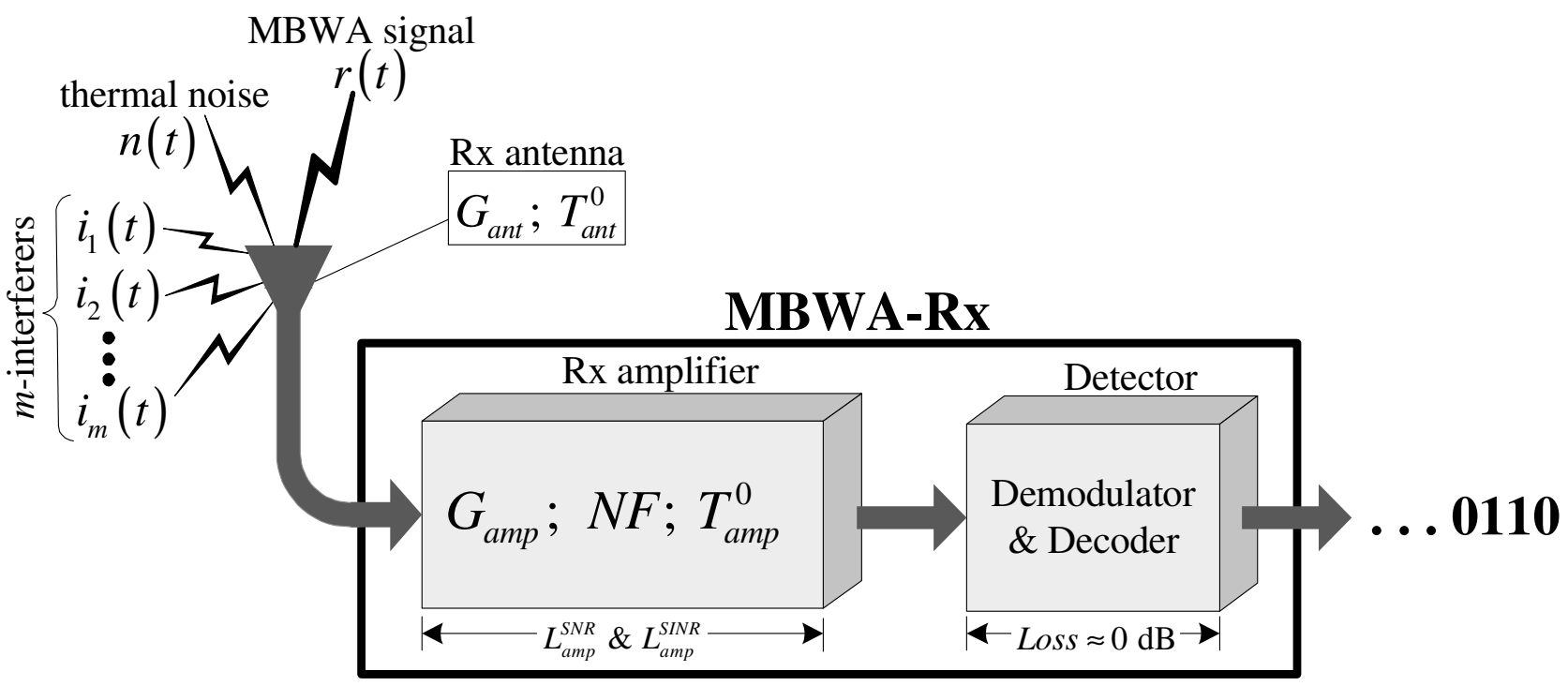

Fig. 1. Model for an MBWA portable receiver

Further, the gain for antenna and amplifier are identified for:

$$\forall G_{ant}, G_{amp} \in \mathbb{R}_{+}^{2} : \exists G_{dB} \in \left\{G_{ant}, G_{amp}\right\} : G_{dB} \geq 0 \tag{1}$$

Indeed, the antenna gain is zero for an idealistic isotropic detector, or a realistic omni-directional radiation. Moreover, a particularly critical parameter is the noise figure $NF_{dB} \in \mathbb{R}$ defined by:

$$NF \triangleq \left. \frac{SNR_{in}}{SNR_{out}} \right|_{T_{ant}^{0}=T_{0}} \tag{2}$$

which is a metric for the amplifier loss at some arbitrary noise temperature $T_0$ perceived at the output of the receiver antenna. And, this value dependent on the input/output signal to noise ratio (SNR), given when:

$$\exists SNR_{in}, SNR_{out} \in \mathbb{R}_{+,*}^{2} : SNR_{in} > SNR_{out} \tag{3}$$

because of the excess noise induced by the amplifier. Also, it is interesting and worthwhile to point out that if the noise temperature for the antenna and the amplifier are governed by:

$$\forall T_{ant}^{0}, T_{amp}^{0} \in \mathbb{R}_{+}^{2} : \exists T_{0} \in \mathbb{R}_{+} : \forall T_{ant}^{0} \approx T_{0} : T_{ant}^{0} < T_{amp}^{0} \tag{4}$$

then it can readily be shown that $NF \gtrsim 3\ \text{dB}$. This observation can serve as a general engineering guideline for design and analysis of such systems.

## 4. Interference Analysis

As it will become evident, different data corruption criteria can be considered as a function of the output signal to interference plus noise ratio (SINR). From its definition, we know that:

$$SINR_{out} = S_{out} / \left(I_{out} + N_{out}\right) \tag{5}$$

where $\exists S_{in}, S_{out} \in \mathbb{R}^2_{+,*}$ and $S_{out} = G_{amp} P_{RX} \geq S_{in}$ are the average MBWA signal power, such that $P_{RX} = \mathrm{E}\left[ r(t)^2 \right]$ is the received power. Further, the aggregate interference power observed by the system can be described as:

$$\exists I_j, I_{agg}, I_{out} \in \mathbb{R}^3_{+,*} : I_{out} = G_{amp} \sum_{j=1}^{m} I_j \geq I_{agg} \tag{6}$$

given that $I_j = \mathrm{E}\left[ i_j(t)^2 \right]$ is the contribution from the $j$-th interfering signal. And for the simplest case, $N_{out}$ represents the thermal noise power modeled by:

$$N_{out} = G_{amp} N_0 \geq \left( N_{ant} + N_{amp} \right) \tag{7}$$

such that $N_0$ is the compound noise of the handheld unit, and $N_{ant}$ and $N_{amp}$ are the corresponding average levels from $n_{ant}(t)$ and $n_{amp}(t)$. Pursuing this further, it can be shown that:

$$\forall T^0_{ant} \approx T_0, NF \gtrsim 3 \text{ dB} : \exists N_{ant}, N_{amp}, N_{out} \in \mathbb{R}^3_{+,*} : N_{ant} < N_{amp} < N_{out} \tag{8}$$

At present, using the above, we can converge to the expression in (9), where $k \approx 1.38065 \times 10^{-23}$ [W/K-Hz] is Boltzmann's constant and $B_{CH} \in \mathbb{R}^*_+$ is the channel bandwidth.

$$SINR_{out} = P_{RX} \Big/ \left\{ I_{agg} + kB_{CH} \left( T^0_{ant} + T^0_{amp} \right) \right\} \tag{9}$$

Admittedly, there are numerous ways to assess the fidelity of a system. For instance, we could quantify the decibel loss based on SNR or SINR, or perhaps a combination of both. No matter the adopted metric, these attributes can generally be determined as a function of the above declaration. For the sake of comparison, in Table 1 the versatility and semantic of the measures are explicitly derived. Within this table, the following indices $L^{SNR}_{RX}, L^{SINR}_{RX}, d \in \mathbb{R}^3_{+,*}$ respectively represent in linear notation: the noise-based receiver loss, noise-interference dependent loss, and system degradation, also known as noise rise. Moreover, in the derivation of these expressions, we assumed an ideal detector, thus no extra loss was taken into account from the demodulator and the decoder.

Despite the multiple means for assessing corruption, degradation is usually specified by protocols, and defined in literature (e.g. [2]) as the ratio of the output SNR to SINR, which is partially dependent on $L^{SNR}_{RX}$ and $L^{SINR}_{RX}$. Thus, from Table 1, we can rewrite it as:

$$d = 1 + I_{agg} \big/ N_0 \tag{10}$$

Table 1. Various system merits in presence of information corruption

| | | | |
|---|---|---|---|
| ***Ratios*** | $SNR_{in} = \lim\limits_{\substack{I_{agg} \to 0 \\ T^0_{amp} \to 0}} SINR_{out}$ | $SNR_{out} = \lim\limits_{I_{agg} \to 0} SINR_{out}$ | $SINR_{in} = \lim\limits_{T^0_{amp} \to 0} SINR_{out}$ |
| ***Losses*** | $L^{SNR}_{RX} \triangleq SNR_{in} / SNR_{out} = 1 + \dfrac{T^0_{amp}}{T^0_{ant}}$ | $L^{SINR}_{RX} \triangleq SINR_{in} / SINR_{out} = 1 + \dfrac{T^0_{amp}}{\left( T^0_{ant} + I_{agg} / kB_{CH} \right)}$ | $d \triangleq SNR_{out} / SINR_{out} = 1 + \dfrac{I_{agg}}{kB_{CH} \left( T^0_{ant} + T^0_{amp} \right)}$ |

Now, if we isolate $I_{agg}$, and since interference, degradation and noise factor are normally provided in decibel nomenclature, then we obtain:

$$\begin{aligned} I_{agg}\left[\text{dBW}\right] &= 10\log_{10}\left\{kB_{CH}\left(T_{ant}^{0}+T_{amp}^{0}\right)\left(d-1\right)\right\} \\ &= 10\log_{10}\left[kB_{CH}\left\{T_{ant}^{0}+T_{0}\left(10^{NF_{dB}/10}-1\right)\right\}\right]+10\log_{10}\left(10^{d_{dB}/10}-1\right) \end{aligned} \tag{11}$$

Certainly, one of the fundamental features of IEEE 802.20 is its support for elevated mobility. And we know that the system throughput and the displacement rate are inversely proportional; therefore, reflecting this interconnection into (11) becomes essential. The MBWA system requirements document [8] identified the minimum spectral efficiency which can be represented by $\eta\left(M,L\right)\in\mathbb{R}_{+}^{*}$ where $M=\left\{p_{M},hs_{M}\right\}$ is the sample space for mobility with elements representing rates for common pedestrian (~3 km/hr) and high-speed automobile (~120 km/hr) events; and $L=\left\{d_{L},u_{L}\right\}$ signifies the link direction for forward and reverse channels. Also, the least peak bit rate per user is a function of the bandwidth, the link, and the extremity described by $R_{b}\left(B_{CH},L,E\right)\in\mathbb{R}_{+}^{*}$ where $E=\left\{l_{E},h_{E}\right\}$ indicates the lower and higher deliverable throughput offered by the standard. In fact, this rate can be shown to equal $B_{CH}R_{0}\left(L,E\right)/B_{0}$ such that $R_0(L,E)$ is a reference data rate with operation band $B_0$. Next, since $\eta=R_{b}/B_{CH}$, after plugging the various interdependencies we get:

$$\eta\left(M,L\right)\leq\eta=R_{b}/B_{CH}\geq R_{b}\left(B_{CH},L,E\right)/B_{CH} \tag{12}$$

This means that the reference and channel bandwidth are respectively proportional to the $R_{0}\left(L,E\right)/\eta\left(M,L\right)$ and $R_{b}\left(B_{CH},L,E\right)/\eta\left(M,L\right)$ ratios.

At this point, as a consequence of (13), we realize that considering the maximum degradation authorized by the standard attains the uttermost tolerance for the interference level.

$$d^{\max}=\sup_{I_{agg}\in\mathbb{R}_{+}^{*}}d\left(I_{agg}\right)=SNR_{out}\Big/\inf_{I_{agg}\in\mathbb{R}_{+}^{*}}SINR_{out}\left(I_{agg}\right)=d\left(I_{agg}^{\max}>I_{agg}\right) \tag{13}$$

Conveying this and the prior steps together, results in the totality of the interference model in dBmW, where the bandwidth and information rate are respectively expressed in units of MHz and Mbps:

$$\begin{aligned} I_{agg}^{\max}\left[\text{dBmW}\right] &\approx -138.6-10\log_{10}\eta\left(M,L\right)+10\log_{10}R_{b}\left(B_{CH},L,E\right) \\ &\quad +10\log_{10}\left(10^{d_{dB}^{\max}/10}-1\right)+10\log_{10}\left\{T_{ant}^{0}+T_{0}\left(10^{NF_{dB}/10}-1\right)\right\} \end{aligned} \tag{14}$$

## 5. Numerical Evaluation and Discussions

Generally speaking, the 802.20 technology allows two possible modes of operations: wideband orthogonal frequency division multiple access (OFDMA) and 625 kHz multicarrier (625k-MC). And, a particular cell network may choose to carry one or both of the modes simultaneously.

Table 2. Numerical parameters

| *Mode* | *Duplexing* | $B_{CH}$ [MHz] | $B_0$ [MHz] | $L$ | $R_0$ [Mbps] |
|---|---|---|---|---|---|
| *OFDMA* | *FDD* | $2.5 \sim 20$ | 1.25 | $d_L$ | $1 \sim 4.5$ |
| | *TDD* | $5.0 \sim 40$ | 2.5 | $u_L$ | $0.3 \sim 2.25$ |
| $625k-MC$ | *TDD* | $0.625 \times n$ $(n \in \mathbb{N}^*)$ | 0.625 / carrier | $d_L$ | 1.493 |
| | | | | $u_L$ | 0.5712 |

| *M* | *L* | $\eta$ [bps/Hz] | *losses* | | *noise temp.* | |
|---|---|---|---|---|---|---|
| $p_M$ (~ 3 km/h) | $d_L$ | 2.00 | $d^{\max}$ | 0.5 dB | $T_0$ | 290 K (17°C) |
| | $u_L$ | 1.00 | | | | |
| $hs_M$ (~ 120 km/h) | $d_L$ | 1.50 | $NF$ | 10.0 dB | $T_{ant}^0$ | 288 K (15°C) |
| | $u_L$ | 0.75 | | | | |

On one hand, the OFDMA scheme supports channels from 2.5 to 20 MHz for frequency division duplex (FDD), and 5 to 40 MHz for time division duplex (TDD) [1]. On the other hand, the 625k-MC uses a bandwidth of 625 kHz per carrier for DL and UL with only TDD, where a single user may utilize multiple carriers [7]. In Table 2, the actual anticipated parameters for this technology are outlined. And within this table, $n$ represents the number of carriers per mobile operator.

Next, applying the reference features from the above table, we derived in Fig. 2 and Fig. 3 the data rates over different operation: modes, bands and conditions. From the plots, we notice that the bit rate fluctuation for DL and UL are nearly 6.53 and 8.75 dB-Mbps for the wideband mode. Thus, greater throughput flexibility is evident in the reverse direction. Whereas for 625k-MC, the rates do not alternate; and the proportion of symbols transmitted from the BS is consistently superior to that of the mobile by roughly 4.17 dB-Mbps for all enabled bands.

As for the interference, we may characterize the results using realistic and practical system parameters in order to derive the desired demarcations [1, 8, 9]. Here, we will start by showing this for the OFDMA mode, and then extend the principle to the 625k-MC scheme.

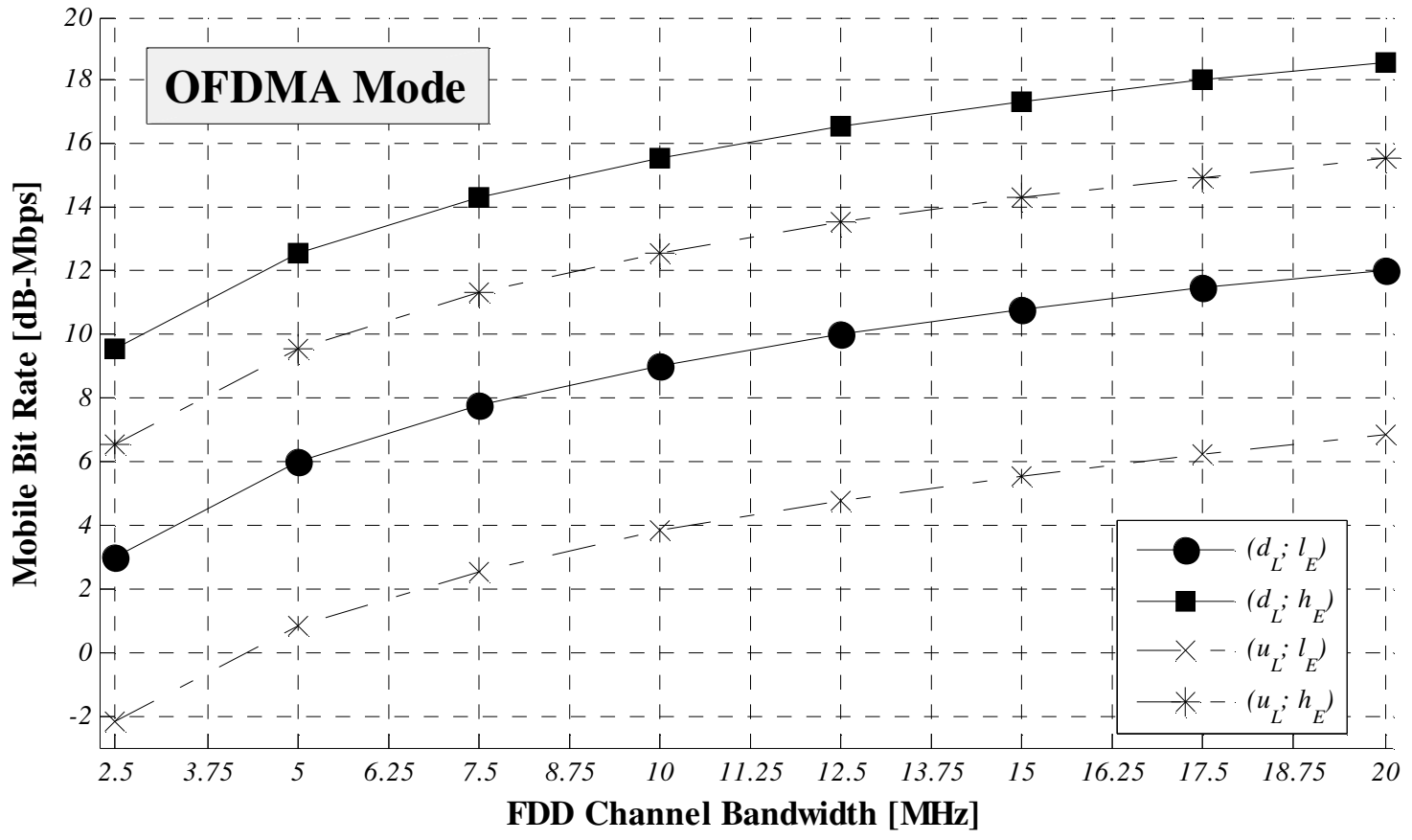

Fig. 2. Wideband peak data rate per user over various link conditions

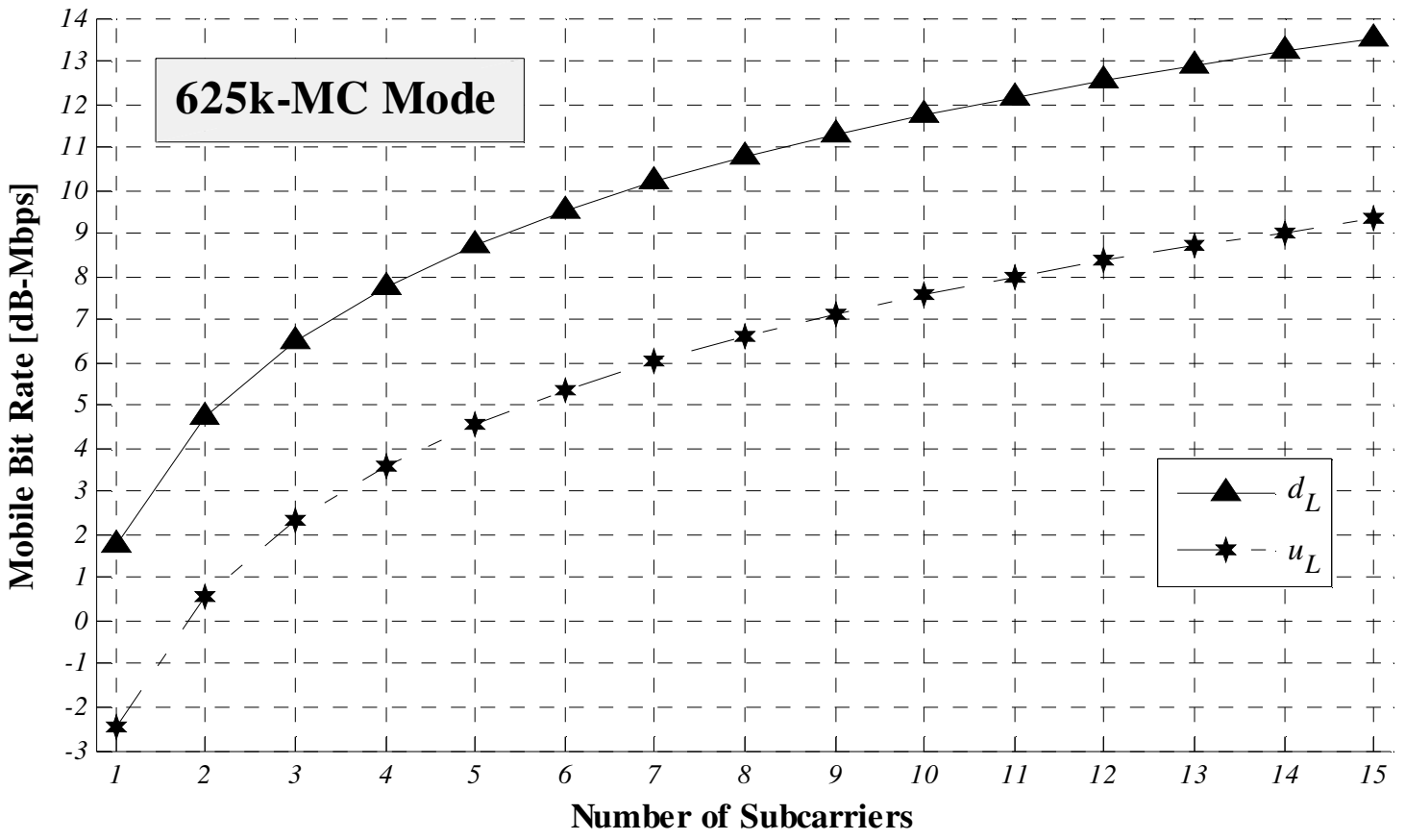

Fig. 3. Multicarrier peak data rate per user for each radiation direction

In fact, using (14), alongside Table 2, and Fig. 2, we obtain the aggregate interference benchmark shown in Fig. 4 and Fig. 5 for pedestrian and high-speed motion. As it can be observed, irrespective of the node velocity, the supreme reference line for the forward and reverse link perfectly coincides.

In addition, it is critical to note that the actual reference boundary for a specific victim system, which is a function of mobility, link, channel, rate, and system factors, will be somewhere within the UL or DL diversity margins shown in the above figures. In fact, once this level is established for a mobile active in a particular environment, if the entire interference power detected by the MBWA device is below the limit, then reliable communication is possible; otherwise an adequate link based on available system parameters is unlikely. In case the latter occurs, as a tradeoff to interference, system designers may for instance opt to augment the transmission power.

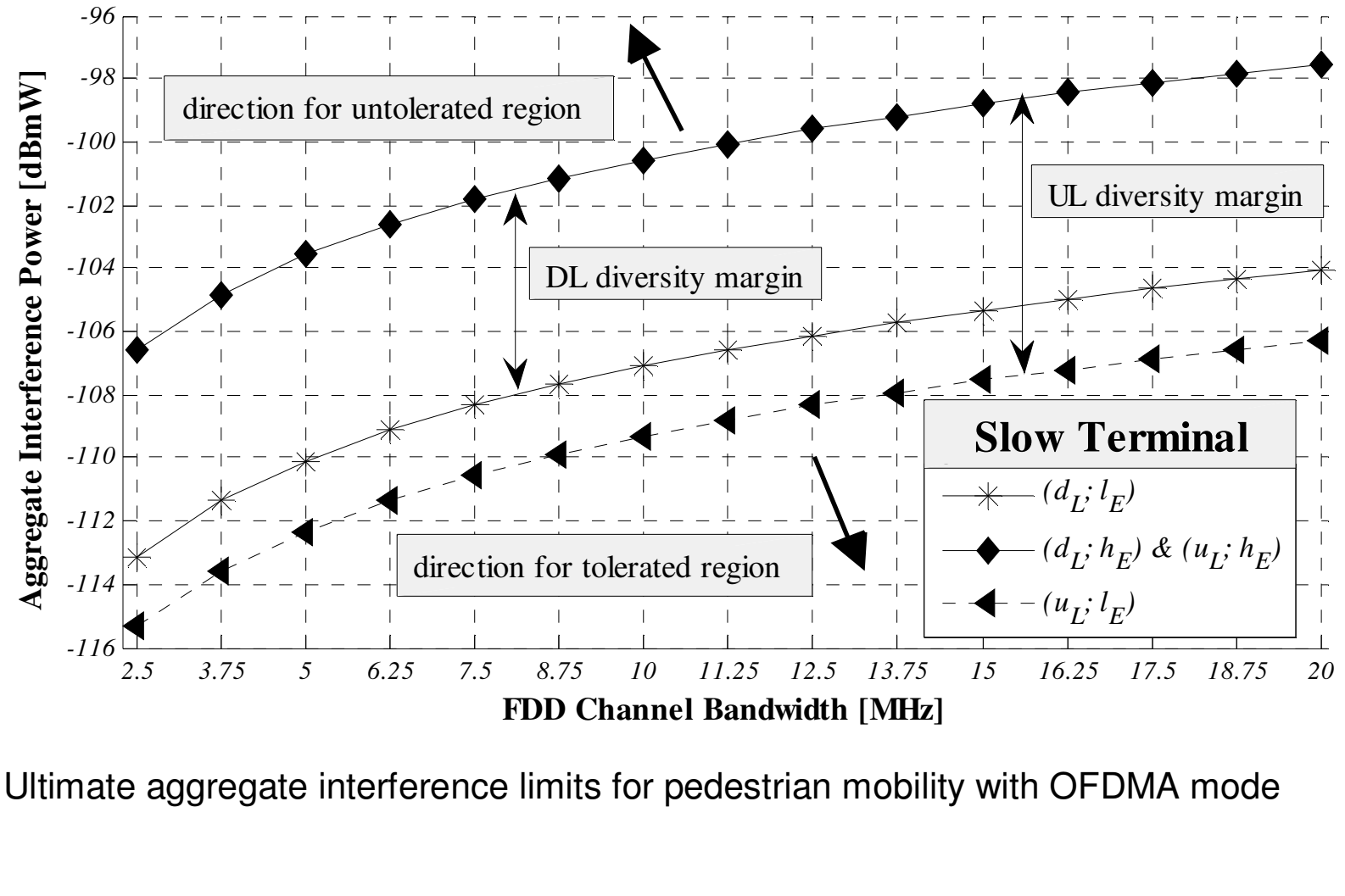

Fig. 4. Ultimate aggregate interference limits for pedestrian mobility with OFDMA mode

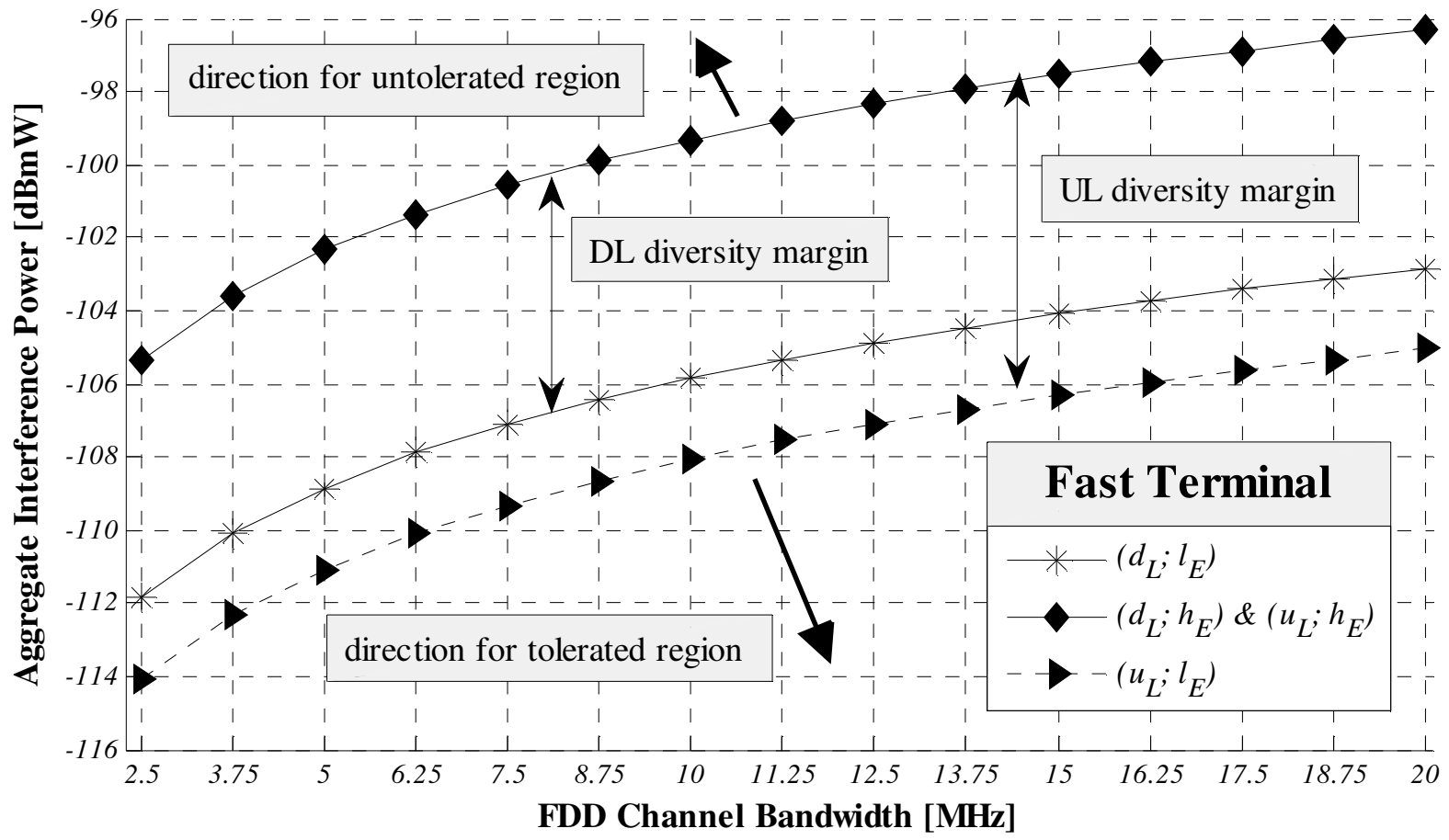

Fig. 5. Ultimate aggregate interference limits for high-speed movement with OFDMA mode

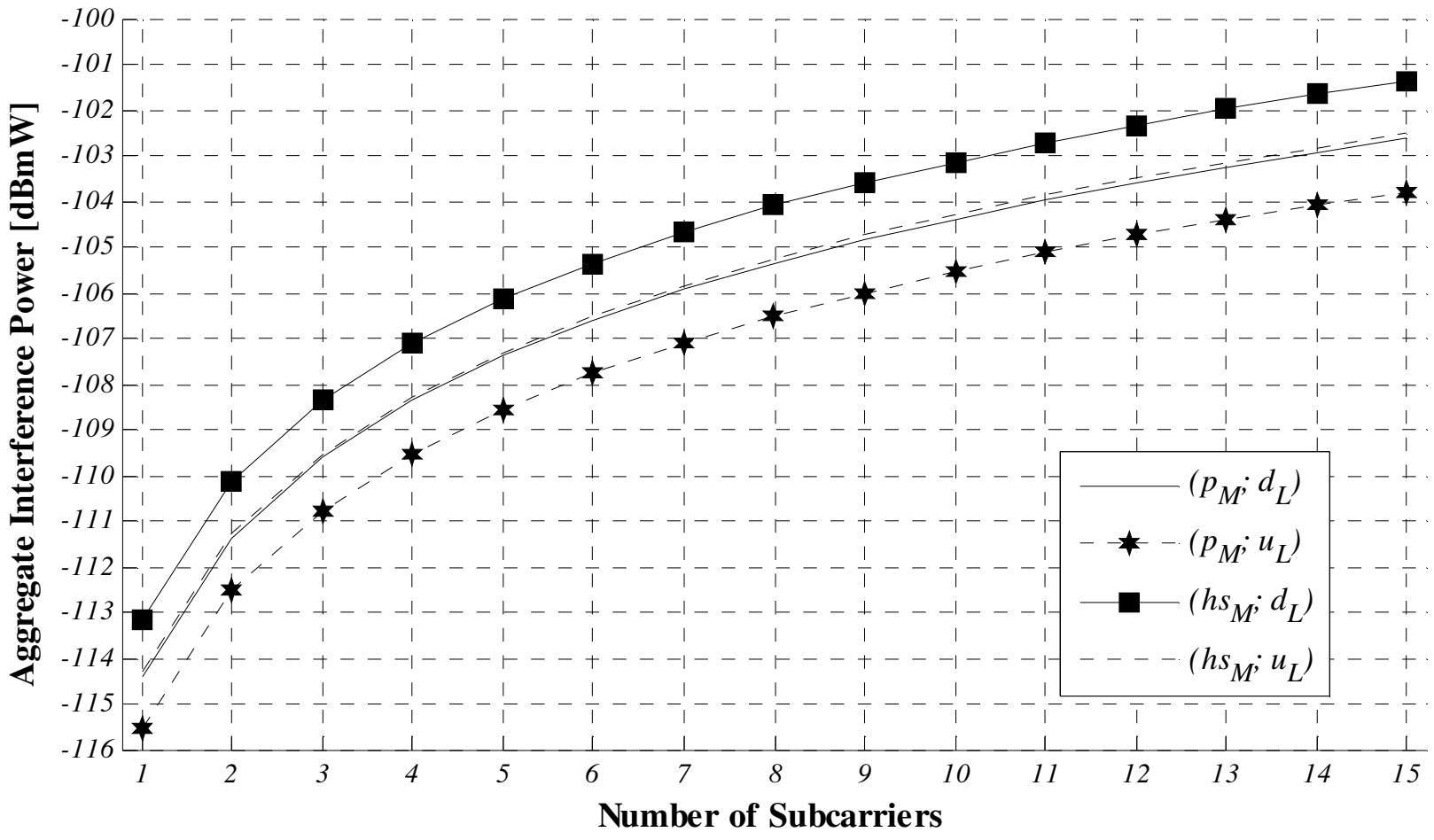

Fig. 6. Utmost interference threshold with the 625k-MC scheme for slow and fast mobiles

In Fig. 6, we also obtain a coexistence benchmark for the multicarrier realization. Interestingly, the baseline for fast units in UL is somewhat greater than slower devices in DL by 0.087 dBmW. Overall, no matter the mode, we also observe that high mobility allows on average 1.33 mW of interference surplus.

## 6. Conclusion

IEEE 802.20 has promising potential for providers and consumers alike because of mobility, spectral efficiency, low latency, long range, and it is specifically optimized for mobile internet protocol (IP) connection and voice over IP applications. Thus, understanding and quantifying its limitations before actual deployment are both fundamental and necessary. The objective of this article was to slightly move in this direction by providing a canonical reference limit for the greatest interference power such that high-fidelity transmission is ensured. In fact, our derivation draws inspiration from previously used demarcations for immobile units, while being exclusively customized for the MBWA technology and all of its strict requirements.

In short, the benchmark derived will be useful as a feasible interference-based quality indicator for system architects and planners during preliminary analysis and design of an IEEE 802.20 system prior to the physical installation of the mobile network. In particular, it can be used as a mechanism to assess the impact of undesirable radios (e.g. a nearby network of high-rate UWB nodes) that overlap the allocated spectrum of the MBWA standard. In fact, this can be emulated via a software subroutine by geometrically positioning interferers (which can be done deterministically or randomly), and verifying the impact that these unwanted nodes have on a reference 802.20 terminal.

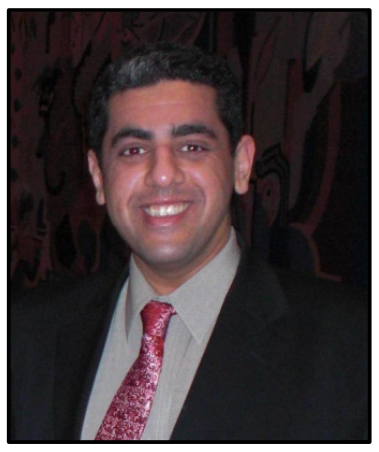

**Mouhamed Abdulla**

He is currently a Systems Engineering Researcher with the Department of Electrical and Computer Engineering at Concordia University in Montréal, Québec, Canada. He received, respectively in 2002 and 2005, a B.Eng. (with Distinction) in Electrical Engineering and an M.Eng. in Aerospace Engineering; currently, he is a Ph.D. Candidate in Telecommunications all at Concordia University. Moreover, for nearly 7 years since 2003, he worked at IBM Canada Ltd. as a Senior Technical Specialist. Mr. Abdulla holds a dozen awards and honors from academia, government and industry; among them the Golden Key Outstanding Scholastic Achievement and Excellence Award (2009) and the IBM Innovation Award (2007). He is professionally affiliated with IEEE, IEEE ComSoc, AIAA, and OIQ. Presently, he is an Associate Editor of the IEEE Technology News Publication. Also, he regularly serves as a referee for a number of journal publications such as: IEEE, Springer, and EURASIP; and contributes as an examination item writer for the prominent IEEE/IEEE ComSoc WCET® Certification Program. His fields of interests include: Wireless Communications, Error Control Coding, DSP, and Avionics. Lately, his research is focused on advancing the fundamentals and characteristics of Random Wireless Networks. Moreover, he has a particular interest in philosophical factors and social aspects related to Engineering Education, and Research Innovation. Recently, his biography was selected for inclusion in the distinguished Marquis Who's Who in the World 2011 publication (28th Edition).

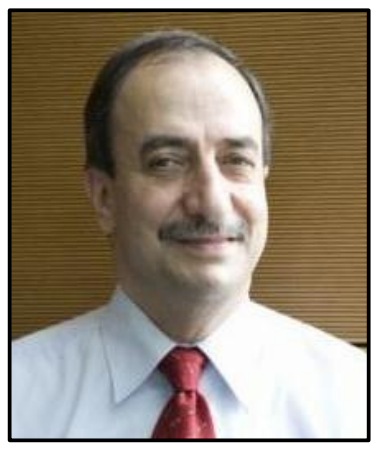

**Yousef R. Shayan**

He received his Ph.D. degree in Electrical Engineering from Concordia University in 1990. Since 1988, he has worked in several wireless communication companies at different capacities. He has been in the R&D departments of SR Telecom, Spar Aerospace, Harris and BroadTel Communications, a company he co-founded. In 2001, Dr. Shayan joined the department of Electrical and Computer Engineering (ECE) of Concordia University as Associate Professor. Since then he has been Graduate Program Director, Associate Chair and Department Chair. Dr. Shayan is founder of the Wireless Design Laboratory, which was established in 2006 based on a major CFI Grant with state-of-the-art equipments for the development of practical systems. In 2008, Dr. Shayan was promoted to the rank of Professor, and was recipient of the Teaching Excellence Award for academic year 2007-2008 awarded by Faculty of Engineering and Computer Science. Since 1985, he has contributed as a Technical Program Committee member for a number of major IEEE conferences. In 2004, he was elected as IEEE Chair for the communications chapter of Montreal, and in 2007 and 2008 was the Treasurer. Lately, in 2010, Dr. Shayan was the Chair of the $23^{rd}$ IEEE CCECE Communications and Networking Symposium. His fields of interests include Wireless Communications, Error Control Coding and Modulation Techniques.